\documentclass{mn2e}
\usepackage{epsf}

\title{The origin of radio haloes and non-thermal emission in 
clusters of galaxies}
 
\author[H.~Liang et al.]
{H.~Liang$^1$, V.~A.~Dogiel$^2$, and M.~Birkinshaw$^{1}$ \\
$^1$ Physics Dept., University of Bristol, Tyndall Avenue, 
Bristol BS8 1TL, England\\
$^2$ P. N. Lebedev Physical Institute, 117924 Moscow, Russia
}
\date{}
 
\pagerange{\pageref{firstpage}--\pageref{lastpage}}

\pubyear{}
 
\begin{document}

\maketitle
 
\label{firstpage}
 
\begin{abstract}
We study the origin of the non-thermal emission from the intracluster
medium, including the excess hard X-ray emission and cluster-wide
radio haloes, through fitting two representative models to the Coma
cluster. If the synchrotron emitting relativistic electrons are
accelerated {\it in situ} from the vast pool of thermal electrons,
then a quasi-stationary solution of the kinetic equation with particle
acceleration through turbulence at high energies ($>200$\,keV)
naturally produces a population of supra-thermal electrons responsible
for the excess hard X-ray emission through bremsstrahlung. Inverse
Compton scattering is negligible at hard X-ray energies in this
case. The radio halo flux density constrains the magnetic field
strength to a value close to that of equipartition $\sim 1
\mu$G. Alternatively, if the relativistic electrons are injected from
numerous localised `external' sources then the hard X-rays are best
explained by inverse Compton scattering from GeV electrons, and
little of the hard X-radiation has a bremsstrahlung origin. In this
case, the magnetic field strength is constrained to $\sim
0.1-0.2\,\mu$G.  Both models assume that the non-thermal emissions are
generated by a single electron spectrum, so that only two free
parameters, well constrained by the observed hard X-ray and radio halo
spectra, are needed in either case. Measurements of the cluster
magnetic field will distinguish between the models.

\end{abstract}
 

\section{Introduction}
 
Since the late 1960s, clusters of galaxies have been known to be strong 
X-ray sources where the thermal  emission from the 
intracluster medium (ICM) produces a diffuse X-ray halo. The ICM is 
diffuse (central electron density of $\sim 10^{-3}$\,cm$^{-3}$) with 
typical temperature $2\times 10^{7}$ to $2 \times 10^{8}$\,K. In 
the 1970s, similarly diffuse radio emission was found in the Coma 
cluster. This radio emission was found to have a steep power law 
index ($\alpha \sim 1$ for $S_{\nu}\propto \nu^{-\alpha}$) which 
indicates a non-thermal origin. Despite efforts to detect similar 
radio haloes in other clusters, until recently few were found with 
diffuse radio emission (see review by Feretti \& Giovannini 1996). 
This led to the belief that radio haloes are rare. 
However, the number of clusters found to have diffuse radio emission 
has more than doubled over the last couple of years, mainly because of 
the availability of moderate brightness sensitivity surveys at 1.4\,GHz, 
e.g. the NVSS \cite{giov99}, as well as high brightness sensitivity 
observations of hot clusters to detect the Sunyaev-Zel'dovich effect 
(e.g. Herbig \& Birkinshaw 1994; Liang et al. 2000).

Diffuse cluster radio emission is now divided into two distinct
classes: haloes and relics. Haloes permeate the entire cluster and
resemble the thermal X-ray emission, but relics are on the cluster
periphery. Both haloes and relics exhibit steep spectral indices
($\alpha \sim 1$) and are extended over $\sim 1$\,Mpc with the haloes
having lower surface brightness. Emission from radio relics is
strongly polarised (up to $\sim 20$\%) but no polarised emission has
been detected in radio haloes.  Radio haloes and relics are believed to
emit by the synchrotron process.  The non-detection of polarised
emission from radio haloes is attributed to  low surface
brightness and to Faraday depolarisation effects due to the mixing of
thermal and relativistic plasma, especially towards the cluster centres. 

However, synchrotron emission from haloes requires a population of
relativistic (GeV energy range) electrons and a cluster-wide magnetic
field. The origins of both particles and field have been a subject of
much debate since the late 1970s (e.g.  Jaffe 1977; Dennison 1980;
Roland 1981). Recent evidence on this question has come through the
observation that radio haloes are preferentially found in high X-ray
luminosity clusters (Giovannini et al. 1999), and that there may be a
steep correlation between the radio power of haloes and the ICM
temperature (e.g. Liang et al. 2000). No upper limits so far
contradict this correlation in the temperature range 5--15\,keV (Liang
2000). The relativistic particles cannot simply leak out of radio
sources in the cluster, because the diffusion path length ($\sim
10$\,kpc assuming Alfv\'en speed propagation) is short compared to the
size of clusters ($> 1$\,Mpc).  Giovannini \& Feretti (2000) found
that while diffuse radio sources were found in bright X-ray clusters,
none of the 11 clusters selected for the presence of a tailed radio
source within the central 300\,kpc showed any diffuse radio
emission. The presence of diffuse radio emission is therefore more
closely connected with the properties of the thermal plasma in the ICM
than the presence of tailed radio sources which were thought to
provide relativistic particles. This gives observational evidence
that particles escaping from radio sources in the cluster are unlikely
to be accelerated to the relativistic energies required for the radio
halo emission. Instead, we interpret the correlation between halo
radio power and ICM temperature, and the similarity in appearance of
the X-ray and radio haloes, as showing that the relativistic electrons
may be accelerated from the thermal pool. The direct cause of this
acceleration may be the merging of subclusters and turbulence in the
ICM (e.g. Tribble 1993, Brunetti et al. 2001).
  
The idea that the same relativistic electrons generate radio emission
through their synchrotron losses and X-ray or gamma-ray emission by
inverse Compton (IC) scattering was first used by Felten and Morrison
(1966) for the interpretation of the galactic diffuse
emission. Later, Cooke et al. (1978) used it for the interpretation
of the X-ray emission from the lobes of the radio galaxy Centaurus A,
assuming that the electrons are scattered by the cosmic microwave
background (CMB) photons.  A similar model was used by Rephaeli (1979)
for the interpretation of the hard X-ray emission from the Coma
cluster: the relativistic electrons responsible for the radio halo
emission can scatter the CMB photons up to X-ray energies through IC
scattering. Efforts have been made to detect this excess of
non-thermal emission above the thermal emission from the hot ICM. It
is easiest to detect at hard X-ray energies (above
$20$\,keV) where the thermal spectrum is expected to cut off
exponentially. Recent observations from {\it Beppo-SAX} have shown
that such a hard X-ray excess does exist in some clusters, e.g. Coma
and A2256 (Fusco-Femiano et al. 1999, 2000).
  
However, alternative theories of the origin of a hard excess have been
proposed by various authors.  En{\ss}lin et al.  (1999) first
suggested that the hard X-ray excess observed in the Coma cluster may
be the result of bremsstrahlung of supra-thermal electrons in the ICM
accelerated by turbulence. Both En{\ss}lin et al. (1999) and Sarazin
\& Kempner (2000) have assumed the acceleration process works on particles with energies of a 
few times $kT_{x}$ (i.e. a few tens of keV), and that the electron energy
distribution has a sharp transition from a Maxwellian to a power law
at a few times $kT_{x}$. Dogiel (2000) pointed out that the
acceleration of particles is only efficient above $\sim 900$\,keV
(i.e. the power law distribution starts at $\sim 900$\,keV), but the
distortion of the Maxwellian is already significant at much lower
energies (as low as $\sim 30$\,keV) due to the flux of thermal
particles running away into the region of acceleration (for details
see Gurevich 1960). It is the `quasi-thermal' particles just above
$\sim 30$\,keV that produce the observed hard X-ray excess through
bremsstrahlung emission.  However, our analysis of the {\it in
situ} acceleration model given below (Sec. 6) shows that the spectrum
of MHD turbulence needed to produce the radio emitting electrons is
very steep which may make this model problematic.

On the other hand, a model where the X-ray and radio emitting
particles are generated by electrons from an `external' injected
population cannot be ruled out. A possible model could be that
particles in the intracluster medium are accelerated only in regions
with high level of turbulence, for example near collisionless bow
shocks generated by motions of hypothetical dark matter haloes (e.g.,
Bykov et al. 2000b).  Radio observations have not shown significant
spatial fluctuations of intensity in radio haloes (e.g. Giovannini et
al. 1993, Liang et al. 2000) which means that we must have either {\it
in situ} acceleration of these electrons everywhere in the
intracluster medium or injection by numerous sources whose separation
is smaller than the path length of the electrons ($\sim 10$\,kpc) or
the length corresponding to the resolution of the radio maps. In the
case of the Coma cluster which is relatively bright and close by, the
highest resolution VLA map of the radio halo corresponds to a length
scale similar to the diffusion path length of the elecrons.
Therefore, the injection model requires a high density of localised
`external' sources.

Below we investigate both the local and global acceleration models and
determine formally the model parameters required to fit the whole
observed spectrum from radio to X-rays   for
the Coma cluster, since this is the best-studied cluster in all
wave-bands.

Our analysis is performed in the framework of a model where spatial
variations of the model parameters are neglected. There are several
reasons for this: a) observations show more or less uniform
distribution of non-thermal emission in the halo; and b)
we do not have enough information about spatial variations of the
hard X-ray flux from the Coma halo, therefore we derive averaged
parameters of particle acceleration based on the total flux of hard
X-rays.

The cosmological parameters adopted are
$H_{0}=50$\,km\,s$^{-1}$\,Mpc$^{-1}$, $q_{0}=0.5$, so that for the
Coma cluster 1 arcmin is 40\,kpc.

\section{The Coma Cluster}
 
Extensive data from radio to X-ray bands exist for the Coma cluster
($z\sim 0.0232$). The best estimates of the ICM temperature are
$T_{x}=8.21\pm 0.09$\,keV from the {\it Ginga} satellite (Hughes et
al. 1993) and $T_{x}=8.25\pm 0.10$\,keV from the {\it XMM-Newton}
satellite (Arnaud et al. 2001). The new high resolution {\it
XMM-Newton} data have shown that at least the central 1\,Mpc of the
cluster is isothermal.  The gas distribution is well
parameterised by a $\beta$-model (Cavaliere et al. 1979) with electron
density distribution given by
\begin{equation}
n_{e}(r)=n_{0}[1+(r/r_{0})^{2}]^{-3\beta/2}
\label{beta}
\end{equation}
where $\beta=0.75\pm 0.03$, the core radius $r_{0}=10\farcm5\pm
0\farcm 6$ (i.e. $408 h_{50}^{-1}$\,kpc), the emission extends to
$10r_{0}$, and the central electron density $n_{0}=(2.89\pm 0.04)
\times 10^{-3} h_{50}^{1/2}$\,cm$^{-3}$, from the {\it ROSAT} all-sky
survey data (Briel et al. 1992).

An excess of hard X-rays above the thermal component has been been
detected in the 20-80\,keV range by {\em Beppo-SAX} satellite
\cite{fusco99}. From Eq.~\ref{beta}, we estimate the average electron
density and squared electron density within the {\em Beppo-SAX} field
of view of $1\fdg 3$ (i.e. a volume of $V\sim 1.7 \times
10^{75}$\,cm$^{3}$) to be $\bar{n}_{e} \sim 1.23\times
10^{-4}$\,cm$^{-3}$ and $\overline{n_{e}^{2}} \sim 5.22\times
10^{-8}$\,cm$^{-6}$. We adopt these values in the discussions that
follow, including the calculation of Fig.~\ref{loss}~\&~\ref{nE}.
 
Coma-C was the first radio halo detected, and its emission has been
measured between 30\,MHz and 4850\,MHz (Deiss et al. 1997 and
references therein).  The spectrum is consistent with a power law of
index $\alpha \sim 1.25$ between 30 and 1400\,MHz. The measurement at
4850\,MHz is an upper limit and the apparent steepening at 2700\,MHz
has been contested (Deiss et al. 1997). Radio spectral index
measurements for such diffuse sources are difficult and a mismatch of
the beam size or difference in the brightness sensitivity of the
observations can often lead to an apparent steepening of the spectral
index. Accordingly, we will disregard the measurements above
1400\,MHz. The largest extent of the radio halo was found to be
$80\arcmin\times 45\arcmin$ on an Effelsberg map at 1.4\,GHz (Deiss
et al. 1997); by contrast, the 1.4\,GHz VLA image which is not as
sensitive to the large scale emission showed a maximum extent of only
$22\arcmin$ (Kim et al. 1990).

\section{Acceleration processes}
 
In most circumstances, the main source of charged particle
acceleration in a cosmic plasma is collisions of electrons and ions
with magnetic field fluctuations.  This leads to a slow, stochastic,
energy gain. Here we consider magneto-hydrodynamic turbulence in the strong
and weak limits. We can determine which regime is most appropriate in
clusters from the spectral indices of radio haloes, since the spectral
signatures of the two cases  differ.

\subsection{Weak Turbulence} 
If the magnetic turbulence is `weak', i.e. the amplitude of magnetic
field fluctuations, $\delta B$, is much less than the strength of the
large scale magnetic field $B_0$, $\delta B\ll B_0$, then the
interaction between the particles and the fluctuations has a resonance
character, which has been discussed in the context of the interstellar
medium (Berezinskii et al. 1990). The resonance condition is
\begin{equation}
\omega(k)-|k_\parallel|v_\parallel\mp\omega_H=0\,.
\label{reson}
\end{equation}
where the different signs correspond to the different wave
polarisations, $k$ and $\omega$ are the wave-number and the 
wave frequency respectively, $v_\parallel$ is the component of the
particle velocity along the magnetic field, and $\omega_H$ is the
gyro-frequency of a particle with the total energy $E_{{\rm tot}}$ and
the charge $Ze$
\begin{equation}
\omega_H={{ZeB_0c}\over E_{{\rm tot}}}\,.
\end{equation}
 
The resonance condition (Eq.~\ref{reson}) can be written in the form
\begin{equation}
k_\parallel=\mp{{ZeB_0}\over{pc(\mu-\omega(k)/kv)}}\,,
\end{equation}
where $v$ is the particle velocity, and $\mu$ is the cosine of the particle 
pitch angle (i.e. the angle between the particle momentum, $p$, and the 
large scale magnetic field, $B_0$). For Alf\'ven waves with dispersion relation
\begin{equation}
\omega(k)=\pm|k_\parallel|v_A\,,
\end{equation}
where $v_A$ is the Alfv\'en velocity, and for fast particles ($v\gg v_A$) the
resonance condition reduces to
\begin{equation}
k_\parallel=|{{ZeB_0}\over{pc\mu}}|= {1\over {r_L|\mu|}}\,.
\end{equation}
Here $r_L$ is the particle gyro-radius. We see that in general
electrons and protons of the same energy (especially for non-relativistic 
energies) are scattered by waves with different wavelength and polarisation. 
Therefore, in some conditions electrons may be preferentially accelerated. 
This may be the case in the Coma halo; and in 
this respect models of hard X-ray emission from Coma which take into 
account of only the electron acceleration are not implausible (e.g.
 Schlickeiser et al. 1987).
 
The process of acceleration may be described as momentum diffusion
with a diffusion coefficient determined by the spectrum of
magneto-hydrodynamic turbulence. If the spectrum of waves is a
power-law
\begin{equation}
W(k)\propto k^{-q},
\end{equation}
the momentum diffusion coefficient depends on the particle momentum as 
(Berezinskii et al. 1990)
\begin{equation}
a(p)=a_{0} p^q\,.
\label{acc}
\end{equation}

It can be shown that in the simple thick target model
the momentum distribution of accelerated particles is given by (see
Gurevich 1960)
\begin{equation} 
f(p)\propto p^{-(q+1)} \,,
\label{sp_hard2} 
\end{equation} 
where $f(p)$ is the electron number density distribution per momentum
volume interval, so that the total number of particles above a
momentum $p$ is given by
\begin{equation}
N(>p)=\int_{p}^{\infty} 4\pi f(p)p^2dp\,,
\end{equation}
and the value of the index $q$ can be deduced from the spectrum of
non-thermal emission produced by the accelerated electrons.

Below (Sec. 6) we perform the necessary analysis in an attempt to estimate
$a_{0}$ and $q$ and hence to define the mechanism of particle
acceleration in the intracluster medium of the Coma cluster.

\subsection{Strong Turbulence} 
In the case of strong magneto-hydrodynamic turbulence, where the field
fluctuations are of the same order as the mean field, both plasma
components, positive and negative, are accelerated. This is seen, for
example, in the acceleration of particles in molecular clouds (Dogiel
et al. 1987) or acceleration by supersonic turbulence, which
effectively acts as an ensemble of shock waves (Bykov and Toptygin
1993). Stochastic acceleration by strong turbulence is described as
momentum diffusion with diffusion coefficient $a (p)$ of the form
(e.g.  Toptygin 1985)
\begin{equation}
a(p)=a_0 p^2\,,
\end{equation} 
which creates a relatively flat spectrum of accelerated particles
\begin{equation} 
f(p)\propto p^{-3} \,.
\label{s_hard} 
\end{equation}
The interpretation of the hard X-ray emission in Coma in terms of this
kind of acceleration was discussed by Dogiel (2000).
 
A similar acceleration process takes place near shock fronts, where
the magnetic field strength jumps by an amount of the order of the
mean field. However, the spectrum of accelerated particles near a
shock front has
\begin{equation} 
f(p)\propto p^{-4} \,,
\label{sp_hard} 
\end{equation} 
steeper than that of an ensemble of shocks (Eq.~\ref{s_hard}).
This case has been analysed for the Coma cluster by Bykov et
al. (2000b).

\section{Particle Kinetic Equation}
To describe the distribution function in both non-relativistic and
relativistic energy ranges it is more convenient to use a canonical
 momentum variable, $p$, than the particle kinetic energy $E$. The electron
number density distribution per unit energy interval, $F(E)$, is related to
the momentum distribution per unit momentum volume, $f(p)$,  by
\begin{equation}
F(E)= 4 \pi f(p)p^{2} \frac{dp}{dE}\,,
\label{en}
\end{equation}
where
\begin{equation}
{{dp}\over{dE}}={1\over{p\sqrt{m_ec^2kT_{x}}}}
\sqrt{p^2+{{m_ec^2}\over {kT_{x}}}}\,,\end{equation} where $m_e$ is
the electron rest mass, $T_{x}$ is the temperature of the thermal gas,
$k$ is the Boltzman constant, and $p$ is the dimensionless particle
momentum, $p=\hat p/\sqrt{m_ekT_{x}}$ where $\hat p$ is the
dimensional momentum.

The generalised Fokker-Planck approximation to the kinetic equation,
describing the particle distribution function $f$ in the intracluster
medium (see Berezinskii et al. 1990), written in dimensionless
variables is
\begin{eqnarray}
&&\frac{\partial f}{\partial t}+u({\bf r})\frac{\partial f}{\partial {\bf r}}-{p\over 3}\frac{\partial u({\bf r})}{\partial {\bf r}} \frac{\partial f}{\partial p}-Q(p,{\bf r})=
\label{eq_total}
\\
&&\frac{1}{p^{2}}\frac{\partial}
{\partial p}\left[{\cal A}(p)\frac{\partial f(p)}{\partial p}+{\cal B}(p) 
f\right]+\nabla (\kappa ({\bf r},p)\nabla f)\,,
\nonumber
\end{eqnarray}
where $u({\bf r})$ is the particle convection, $\kappa$ is the spatial
diffusion coefficient, and $Q$ describes the distribution of sources
emitting fast particles into the intracluster medium. We define
non-dimensional variables $t=\hat t\cdot \nu_0$, $r=\hat r/L$, where
$\hat t$ and $\hat r$ are the dimensional variables, $L$ is the size
of the cluster, and we define the characteristic frequency $\nu_0$ as
\begin{equation}
\nu_0={{2\pi \bar{n}_{e}c^{2}r_{e}^{2}m_{e}}\over{\sqrt{m_{e}kT_{x}}}}\,,
\end{equation}
$\bar{n}_{e}$ is the average electron number density, and $r_e$ is the
classical electron radius $r_{e}=e^{2}/m_{e}c^{2}$. For the Coma halo of
$\bar{n}_{e}=1.23\times 10^{-4}$\,cm$^{-3}$ within a volume of $1.7\times 10^{75}$\,cm$^{-3}$ and $T_{x}=8.2$\,keV (Sec. 2), $\nu_0\simeq
6.4\times 10^{-17}$ s$^{-1}$.

The frequency $\nu_0$ is proportional to the frequency $\nu_1$ of
thermal particle Coulomb collisions at kinetic energy $kT_{x}$
\begin{equation}
\nu_1=2 \nu_0 \left({{m_{e}c^2}\over{kT_{x}}}\right)\ln\Lambda\,,
\end{equation}
where $\ln\Lambda$ is the Coulomb logarithm.  The
frequency $\nu_1\simeq 5.3\times 10^{-14}$ s$^{-1}$ for the Coma halo.

In Eq.~\ref{eq_total}, ${\cal B}$ describes the particle energy gain
and loss and ${\cal A}$ the momentum diffusion due to Coulomb
collisions with background particles and interactions
with the magnetic fluctuations,
\begin{equation}
{\cal B}(p)=p^{2}\left[-\left(\frac{dp}{dt}\right)_{ion}-
\left(\frac{dp}{dt}\right)_{synIC}-\left(\frac{dp}{dt}\right)_{brem}\right]\,,
\end{equation}
\begin{equation}
{\cal A}(p)=p^{2}\left[-\left(\frac{dp}{dt}\right)_{ion}\frac{\gamma}{\sqrt{\gamma^{2}-1}}\sqrt{\frac{kT_{x}}
{m_{e}c^{2}}}+a_{0} p^{q}\right]\,,
\label{ap}
\end{equation}
where $\gamma=1+E(p)/m_{e}c^{2}$ is the Lorentz factor, and the
dimensionless parameter $a_0$ is related to the dimensional parameter
$\hat a_{0}$ as
\begin{equation}
a_0={{\hat a_{0} (\sqrt{m_ekT_{x}})^{q-2}}\over{\nu_0}}.
\end{equation} 

As one can see from Eq.~\ref{ap} the process of momentum
diffusion is determined by Coulomb collisions with background particles
(the first rhs term of Eq.~\ref{ap}) and collisions with
electromagnetic fluctuations (the second rhs term of Eq~\ref{ap}).

The $dp/dt$ terms are rates of momentum loss due to ionisation,
synchrotron plus inverse Compton and bremsstrahlung emission
respectively (e.g. Hayakawa 1969). The ionisation loss term is
\begin{eqnarray}
&& \left(\frac{dp}{dt}\right)_{ion} = -\frac{1}{p}\sqrt{p^{2}+
\frac{m_{e}c^{2}}{kT_{x}}}\frac{\gamma}{\sqrt{\gamma^{2}-1}}\\
&& \times \left[\log\left(\frac{E(p)m_{e}c^{2}(\gamma^{2}-1)}
{h^{2}\omega_{p}^{2}\gamma^{2}}\right)+0.43\right]\,,
\nonumber
\end{eqnarray}
where the plasma frequency $\omega_{p}=\sqrt{4\pi e^{2}\bar{n}_{e}/m_{e}}$. The combined synchrotron and IC loss term is 
\begin{eqnarray}
&& \left(\frac{dp}{dt}\right)_{synIC} = -\frac{1}{p}\sqrt{p^{2}+
\frac{m_{e}c^{2}}{kT_{x}}} 
\frac{32\pi r_{e}^{2}}{9\sqrt{m_{e}kT_{x}}}\frac{1}{\nu_{0}}\\
&& \times \left[U_{cmb}+U_{mag}\right]\left(\frac{E(p)}{m_{e}c^{2}}\right)^{2}\,,
\nonumber
\end{eqnarray}
where $U_{cmb}$ is the energy density of the CMB, and $U_{mag}$ is the
magnetic energy density. The bremsstrahlung loss term is given by
\begin{eqnarray}
&& \left(\frac{dp}{dt}\right)_{brem} = -\frac{1}{p}\sqrt{p^{2}+
\frac{m_{e}c^{2}}{kT_{x}}} 
\frac{\bar{n}_{e} m_{e}c^{2}e^{2}r_{e}^{2}}{hc\sqrt{m_{e}kT_{x}}} 
\frac{1}{\nu_{0}}\\
&& \times \left[\log\frac{2(E(p)+m_{e}c^{2})}{m_{e}c^{2}}-\frac{1}{3}\right]
\frac{E(p)}{m_{e}c^{2}}\,.
\nonumber
\end{eqnarray}
In comparison to the loss terms, the average rate of acceleration due
to momentum diffusion (Eq.~\ref{acc}) is
\begin{eqnarray}
&& \left(\frac{dp}{dt}\right)_{acc} \sim a_{0}p^{q-1}.
\end{eqnarray}

\begin{figure*}
\epsfxsize 300pt \epsfbox{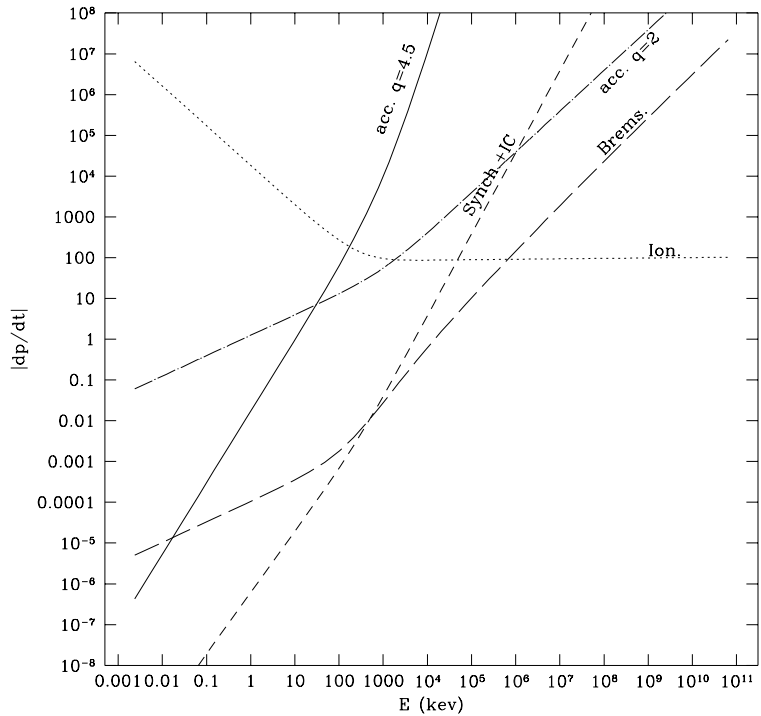}
\caption{The average rate of momentum loss and acceleration as a
function of energy in the Coma cluster ($\bar{n}_{e} \sim 1.23\times
10^{-4}$\,cm$^{-3}$ and $\overline{n_{e}^{2}} \sim 5.22\times
10^{-8}$\,cm$^{-6}$ within the {\it Beppo-SAX} field of view). The
solid curve gives the rate of acceleration, where $a_{0}=0.43$ and
$q=4.5$ corresponding to the best fit to observational data (see
Sec. 6); the dotted curve gives the rate of ionisation loss; the
dashed curve gives the rate of synchrotron and inverse-Compton loss;
the long-dashed curve gives the rate of bremsstrahlung loss. For
comparison with Dogiel (2000), we also show the acceleration rate
corresponding to strong turbulence with $a_{0}=5$ and $q=2$. } \label{loss}
\end{figure*}

These loss terms and the acceleration term are plotted as a function
of particle energy in Fig.~\ref{loss} for the Coma cluster. At low
energies ionisation loss dominates, and at high energies either the
acceleration term or the combined synchrotron and IC term dominates
depending on the value of $q$.

Equation~\ref{eq_total} describes the particle spectrum in the
energy range from thermal energies, where the spectrum is Maxwellian, up
to non-thermal energies,  as well as the particle spatial distribution.
In general it is difficult to obtain a solution,
though in some special cases it is possible to solve it analytically or
numerically (see Bulanov and Dogiel 1979).

Fortunately, in many cases the spatial distribution is unimportant.
 Whether or not a model without spatial terms can be applied to Coma
 is rather questionable.  At the moment we do not have enough
 information to prove or disprove this.  

If we ignore the spatial variation, then we can integrate
Eq.~\ref{eq_total} over the volume of the object and the terms
describing the spatial distribution $f$ converge to a
single term, which can be represented as $f/t_{{\rm eff}}$ where
$t_{{\rm eff}}$ is the effective escape time of the particles due to
their spatial propagation. If the characteristic escape time is much
greater than the energy loss timescale, we can neglect the term
$f/t_{{\rm eff}}$.  Then Eq.~\ref{eq_total}, rewritten for the
spatially-averaged distribution function $f(p)$, can be presented as
\begin{equation}  
\frac{\partial f}{\partial t}-
\frac{1}{p^{2}}\frac{\partial}
{\partial p}\left[{\cal A}(p)\frac{\partial f(p)}{\partial p}+{\cal B}(p) 
f\right]=Q(p)\,.
\label{eq_insitu}
\end{equation}

Two possible fast particle production models are analysed for
the Coma halo in Sec. 6 and 7:
\begin{itemize}
\item {\it in situ} acceleration everywhere in the halo,
where the spectrum of non-thermal particles is established by acceleration
processes represented in Eq.~\ref{eq_insitu} by the momentum
diffusion term ${\cal A}(p)$ and balanced by the loss term, ${\cal B}(p)$; and

\item injection into the intracluster medium from many relatively
small regions where acceleration takes place (perhaps strong shocks,
or regions with high levels of magnetic turbulence). These regions are
supposed to be more or less uniformly distributed in space with no
acceleration between them. The spectrum in the intracluster
medium is then fully determined by energy loss processes and 
the injection spectrum, i.e. ${\cal B}(p)$ and $Q(p)$.
\end{itemize}

\section{Processes of Non-Thermal Emission Generated by Fast Electrons} 

To determine the parameters for both models we need to calculate the
X-ray and radio spectra expected from the non-thermal electrons
and compare our results with observational data. In this section, we
collect the equations describing the radiation processes.

\subsection{Non-thermal Bremsstrahlung Emission}
The bremsstrahlung flux density $F_{brem}$ in units of 
ph~cm$^{-2}$~s$^{-1}$~keV$^{-1}$ is given by
\begin{equation}
F_{brem}=\frac{V}{4\pi d_{L}^{2}} \int_{E_{\gamma}}^{\infty} 
\overline{n_{e}^{2}} \frac{F(E)}{\bar{n}_{e}} \sqrt{\frac{2E}{m_{e}}} 
\frac{d\sigma_{x}}{dE_{\gamma}}\,dE
\end{equation}
where $V$ is the emission volume, $d_{L}$ is the luminosity distance,
$E_{\gamma}$ is the photon energy (not to be confused with the
particle energy $E$), and the emission cross section is given by
\begin{equation}
\frac{d\sigma_{x}}{dE_{\gamma}}=\frac{16}{3}\frac{e^{2}}{\hbar c}
\frac{r_{e}^{2} m_{e}c^{2}}{E E_{\gamma}} 
\log\left(\frac{\sqrt{E}+\sqrt{E-E_{\gamma}}}{\sqrt{E_{\gamma}}}\right)
\end{equation} \cite{hay69}.
Non-thermal bremsstrahlung radiation in the hard X-ray range comes
from electrons in the energy range $\sim 30-100$\,keV.
 
\subsection{Inverse Compton Emission}
Here we calculate the IC flux density $F_{IC}$ in units of
ph~cm$^{-2}$~s$^{-1}$~keV$^{-1}$ due to the scattering of CMB photons
by relativistic particles.  Since we are only interested in the IC
scattering above 100\,eV (i.e. EUV and above), we are interested only
in electrons with $E>10^{5}$\,keV. We will show (Fig.~\ref{nE}) that
the electron energy distribution follows a power law with slope $1-q$
for $E>10^{4}$\,keV.  We can use the analytic solution for a power law
distribution in the Thomson limit ($\sqrt{E_{\gamma}k T_{r}} \ll
m_{e}c^{2}$) given in Blumenthal \& Gould (1970; Eq.~2.65) for this
power law to find
 
\begin{equation}
F_{IC}=\frac{V}{4\pi d_{L}^{2}} \frac{r_{e}^{2} }{\pi h^{3} c^{2}} 
K_{e}(m_{e}c^{2})^{2-q} (kT_{r})^{\frac{q+4}{2}} \zeta(q-1) E_{\gamma}^{-q/2}\,,
\end{equation}
where numerical values of $\zeta(q-1)$ are given in Table~1 of
Blumenthal \& Gould (1970), $T_{r}$ is the temperature of the CMB (not
to be confused with the electron temperature $T_{x}$),
$K_{e}=F(E_{min}) E_{min}/(q-2)$ is the normalisation for the power
law distribution and $E_{min}$ is typically $10^{4}$\,keV. Electrons
of energies $\sim 5$\,GeV inverse Compton scatter photons of the CMB
to hard X-ray energies $\sim 60$\,keV.
 
\subsection{Synchrotron Radio Emission}
We calculate the synchrotron flux density in Jy from the 
electron energy distribution using Ginzburg \& Syrovatskii (1964):
\begin{equation}
F_{radio}=\frac{V}{4\pi d_{L}^{2}} 10^{23} \int^{E_{max}}_{E_{min}} 
\sqrt{3} r_{e} e B F(E) f_{x}\,dE\,,
\end{equation}
where
\begin{equation}
f_{x}=\frac{1}{2}\int^{\pi}_{0}\sin^{2}\theta \frac{\nu}{\nu_{c}} 
\int^{\infty}_{\frac{\nu}{\nu_{c}}} K_{5/3}(\eta) \,d\eta \,d\theta\,,
\end{equation}
$\nu_{c}=4.21B\sin\theta (E/m_{e}c^{2})^{2}$\,MHz, and $B$ is in
units of Gauss.  Since in our case the electron energy distribution is
a power law for $E>10^{4}$\,keV and the synchrotron emission in the
radio frequency range is given by the GeV electrons, we can safely use
the analytic equation for a power law electron energy distribution
given in Eq.~6.10 of Ginzburg \& Syrovatskii (1964). For radio
emission above $\sim 10\,$MHz, we need electrons above energies of
$\sim 1$\,GeV for $B\sim 1\,\mu$G.

\section{Model of {\it in situ} Acceleration}

If the injection term, $Q(p)=0$, then apart from the ionisation and
radiation losses, the spectrum is completely determined by two
processes: Coulomb collisions, which form the thermal and
quasi-thermal part of the spectrum, and stochastic acceleration, which
forms the spectrum at high energies (see Fig.~\ref{loss}).

A quasi-stationary solution to Eq.~\ref{eq_insitu}, in the sense that
the presence of a `weak' acceleration mechanism leads to the
continuous acceleration of particles above a certain injection energy,
is given by Gurevich (1960):
\begin{equation} 
f(p) = \sqrt{\frac{2}{\pi}} \bar{n}_{e} \exp\left(-\!\int^{p}_{0}
\frac{{\cal B }(v)}{{\cal A}(v)}\,dv\right){\cal G}\,,
\label{fp}
\end{equation}
where
\begin{equation}
{\cal G}=\frac{\int^{\infty}_{p}\frac{dv}{{\cal A}(v)}\exp(
\int^{v}_{0}\frac{{\cal B}(t)}{{\cal A}(t)}\,dt)}{\int^{\infty}_{0}
\frac{dv}{A(v)}\exp(\int^{v}_{0}\frac{{\cal B}(t)}{{\cal A}(t)}\,dt)}\,,
\end{equation}

\begin{figure*}
\epsfxsize 300pt \epsfbox{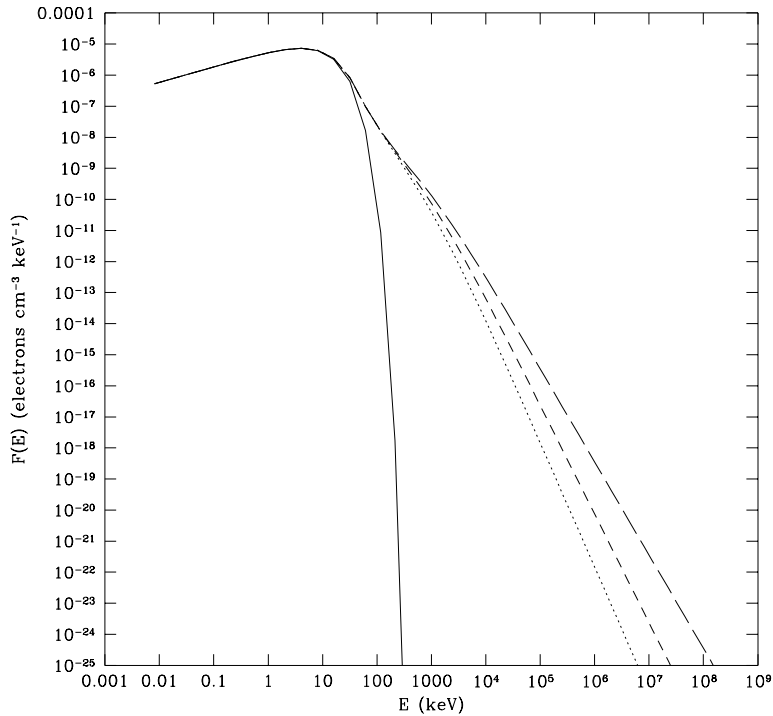}
\caption{Energy distribution of electron density for a pure thermal
distribution (solid curve), and various acceleration parameters
($a_{0}$) and spectral indices $q$. The dotted line is for $a_{0}=0.3$
\& $q=5$; the dashed line is for $a_{0}=0.43$ \& $q=4.5$; and the
long-dashed line is for $a_{0}=0.6$ \& $q=4$. The cluster is assumed
to be isothermal with $T_{x}=8.2$\,keV and average electron density
$n_{e} \sim 1.23\times 10^{-4}$\,cm$^{-3}$ within the {\it Beppo-SAX}
field of view. The electrons of order $\sim 10^{7}$\,keV are
responsible for the radio emission and IC emission in the hard X-ray regime, and the electrons at $\sim 50$\,keV
give rise to non-thermal Bremsstrahlung emission in the hard X-rays.}
\label{nE}
\end{figure*}

The energy distribution of the electron number density  thus
calculated is shown in Fig.~\ref{nE} for 3 sets of ($a_{0}$,$q$).
It can be seen that $F(E)$ starts to deviate from a pure Maxwellian at a
few tens of keV and reaches a power law at  $\sim 1-10$\,MeV.

\begin{figure*}
\epsfxsize 300pt \epsfbox{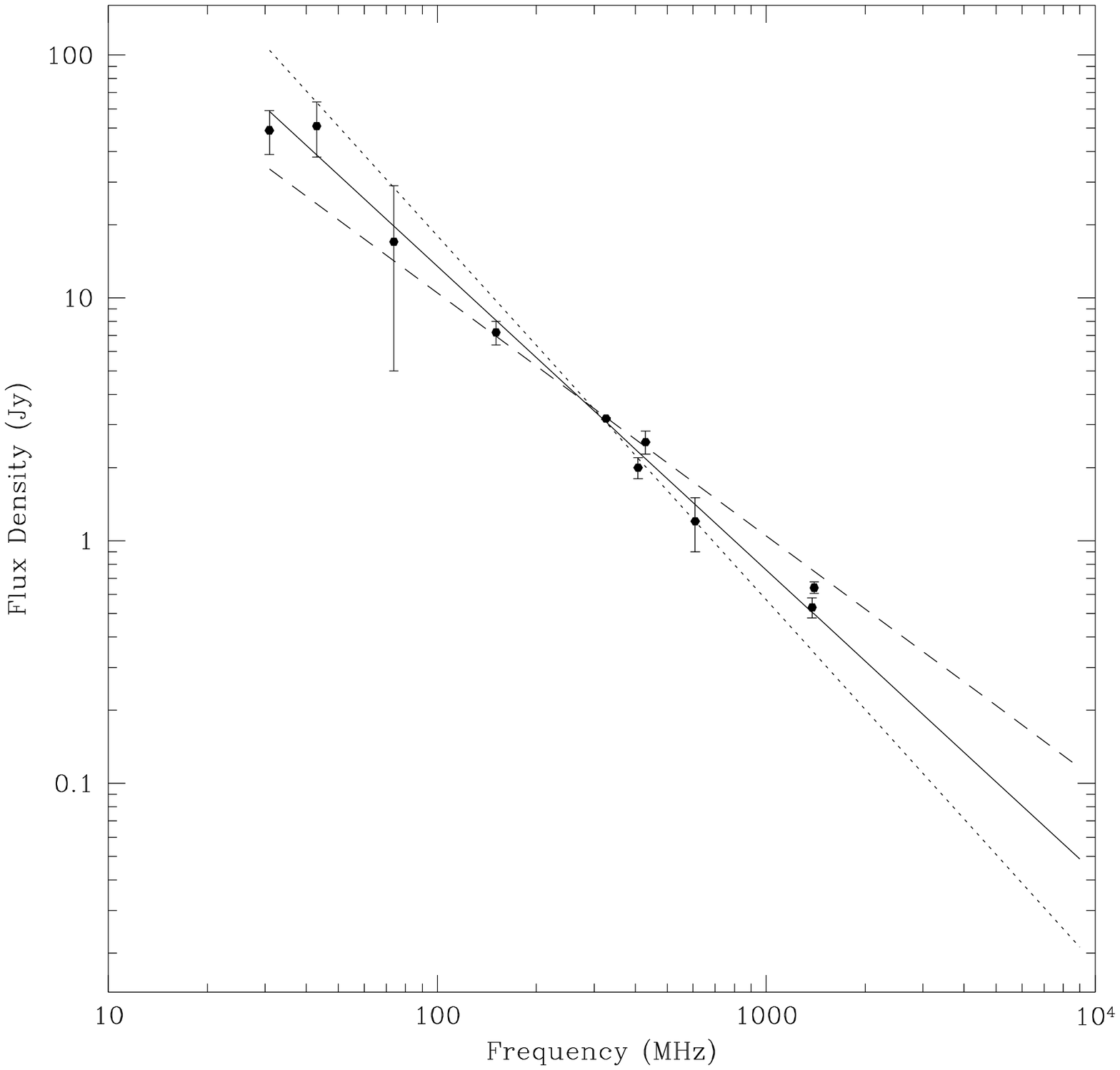}
\caption{Synchrotron flux density derived from the electron energy
  distributions in Fig.~\ref{nE} for the 3 sets of acceleration
  parameters ($a_{0}$,$q$) and their corresponding magnetic field
  $B$. The magnetic field was adjusted such that the model flux
  densities fit the data at 326\,MHz. The observed flux density for the
  Coma radio halo are shown along with the 3 models. The solid line
  corresponds to $q=4.5$ ($B=1.51 \mu$G); the dotted line corresponds to
  $q=5$ ($B=9.7 \mu$G); and the dashed line is for $q=4$ ($B=0.16
  \mu$G). The average density within the radio emitting region (i.e. an
  equivalent radius of $\sim 34\arcmin$) is 
 $\sim 1.4\times 10^{-4}$\,cm$^{-3}$. }
\label{radio}
\end{figure*}

\begin{figure*}
\epsfxsize 300pt \epsfbox{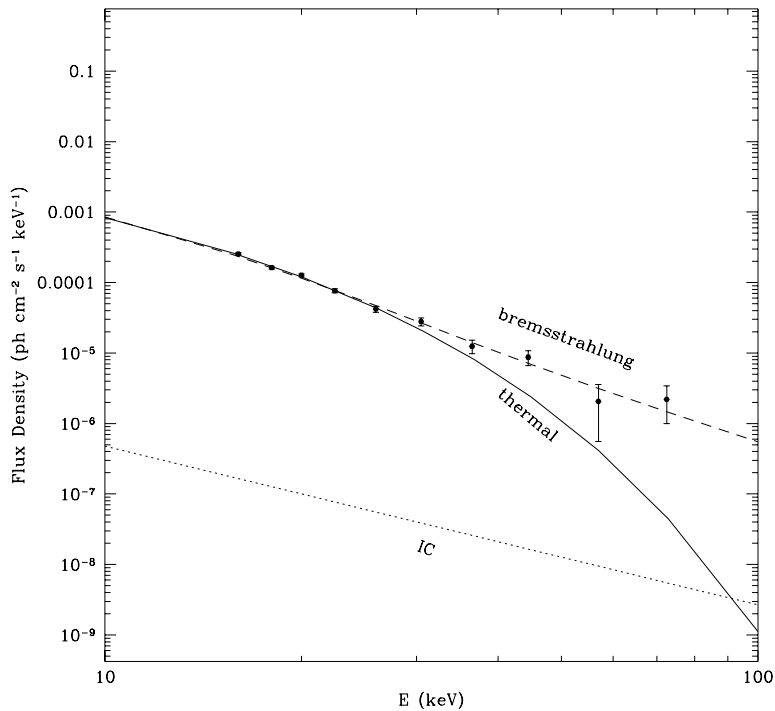}
\caption{Models for hard X-ray (10-100 keV) emission in the Coma
cluster. The solid curve shows a pure thermal bremsstrahlung spectrum
calculated from the Raymond-Smith code (Raymond \& Smith 1977) for
$T_{x}=8.2$\,keV; the dashed curve corresponds to the non-thermal
bremsstrahlung emission derived from one of the electron energy distributions
shown in Fig.~\ref{nE} (dashed line; $a_{0}=0.43$ \& $q=4.5$); the
dotted curve shows the X-ray emission from inverse-Compton scattering
of the CMB by the relativistic electrons of the same electron
distribution. The data points correspond to the observed hard X-ray
flux densities from {\it Beppo-SAX}. }\label{xray2}
\end{figure*}

\begin{figure*}
\epsfxsize 300pt \epsfbox{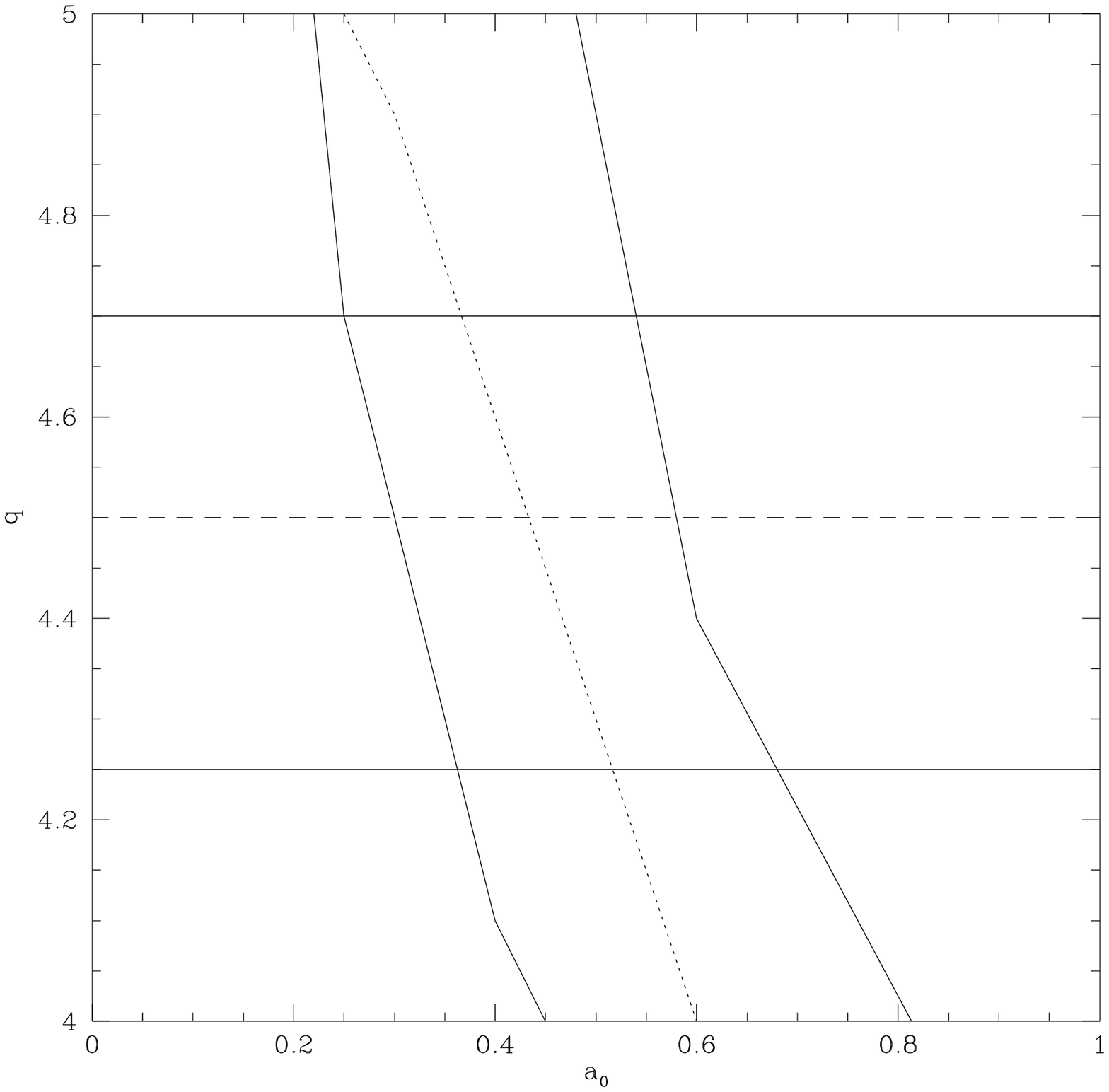}
\caption{The acceleration parameter $a_{0}$ and particle spectral
index $q$ for Coma based on the {\it Beppo-SAX} hard X-ray data and
radio halo spectral index. The permitted ranges are $a_{0} \sim
0.25-0.7$ and $q \sim 4.3-4.7$ with best fit values $a_{0}\sim 0.43$
and $q\sim 4.5$. The dotted curve gives the best fit values of
($a_{0}$,$q$) based on hard X-ray data, and the solid curves give the
95\% confidence range. Similarly, the dashed line gives the best-fit
$q$ value based on the radio halo spectral index, with the solid lines
giving the upper and lower 95\% bounds.} \label{aq}
\end{figure*}

Radio spectra implied by these ($a_{0}$,q) parameter sets are shown in
Fig.~\ref{radio}, and the hard X-ray spectra are shown in
Fig.~\ref{xray2}.  From Eq.~\ref{fp}, it follows that the momentum
spectral index of the accelerated particles is $q-1$, hence $q$ is
related to the the spectral index $\alpha$ of the synchrotron emission
as $q=2\alpha+2$.  The observed radio spectrum (Fig.~\ref{radio})
constrains the electron momentum spectral index to $q\simeq
4.5^{+0.2}_{-0.25}$. Figure~\ref{aq} shows that the best fit
acceleration parameters from the radio and hard X-ray data are
$a_{0}\simeq 0.43^{+0.27}_{-0.18}$, $q\simeq 4.5\mp 0.2$, where the
errors correspond to 95\% confidence. The corresponding magnetic field
derived from adjusting the model flux density to fit the data at
326\,MHz, is $B\simeq 1.51^{+2.31}_{-0.95}\,\mu$G. The magnetic field
is well constrained in this model (with only 2 free
parameters) by the X-ray  and radio data, and is
consistent with the limit deduced from Faraday rotation measurements,
$B\la 5.9 h_{50}^{1/2}\,\mu$G on field tangling scales of $<1$\,kpc
(Feretti et al. 1995).  In comparison, the equipartition magnetic
field is $B\sim 1$\,$\mu$G assuming that there are equal number of
relativistic protons and electrons and that the emission filling factor is
1. Any independent estimates of the cluster global magnetic
field would provide a test for the validity of the model.
 
In the framework of this model the origin of the hard non-thermal
X-rays is bremsstrahlung since the flux of IC X-ray photons is
negligible. Figure~\ref{bric} shows the expected ratio of the
bremsstrahlung to IC flux density for the Coma cluster at 70\,keV as a
function of the momentum diffusion coefficient spectral index, $q$. If
the GeV electrons responsible for the radio halo emission are the
result of {\it in-situ} acceleration, then the observed radio halo
spectrum ($\alpha \sim 1.25$ or $q\sim 4.5$) implies that the
contribution from the bremsstrahlung of 30-100\,keV electrons is
always greater than that of the IC scattering of CMB photons by the
GeV electrons.  This result is independent of the magnetic field,
since neither the bremsstrahlung nor the IC emissivity  depends on
the magnetic field. 

CMB photons can also be up-scattered to the
extreme-ultraviolet (EUV) energy range by electrons of energy
$2.5\times 10^{5}$\,keV through inverse Compton scattering. The
detection of excess EUV emission above that of thermal bremsstrahlung
has been reported in the Coma cluster using data gathered by the
Extreme Ultraviolet Explorer (EUVE; Lieu et al. 1999 and references
therein). However, this result is controversial (see e.g. Bowyer et
al. 1999), hence we have chosen to predict the excess EUV flux
density in the framework of our model instead of fitting it to the
data. We estimate the expected excess EUV flux density at 100\,eV to
be $1.9\times 10^{-3}$\,ph\,cm$^{-2}$\,s$^{-1}$\,keV$^{-1}$ within a
$15^{'}$ radius which is nearly 2 orders of magnitude below the
EUVE sensitivity. However, given the controversy over the
observational data, it can not be used to rule out the model without
further independent measurements. 

It is interesting to note that, for $q<4$, IC starts to dominate over
bremsstrahlung in the hard X-ray regime, but for $q\la 2$
bremsstrahlung dominates again because synchrotron and IC losses
exceed the gains from acceleration and causes the electron spectrum to
cut off at $\sim 1$\,GeV. To produce X-ray emission above 20\,keV
through IC scattering of the CMB, we need electrons with energies
above $\sim 3.5$\,GeV.  For example, if the radiating particles are
accelerated by strong turbulence, that is $q=2$ (hence $a_{0}=5$ to
fit the hard X-ray data assuming a bremsstrahlung origin), then
ionisation losses dominate over acceleration below 1\,MeV (see
Fig.~\ref{loss}). The gain from acceleration overwhelms particle
energy losses in the energy range of 1\,MeV - 1\,GeV. Hence, the
maximum energy of accelerated particles is $\sim 1$\,GeV. At higher
energies, the combined synchrotron and IC losses dominates over
acceleration (see Fig.~\ref{loss}), hence strong turbulence is unable
to produce radio emitting GeV electrons. Acceleration by strong
turbulence can produce only relatively low energy electrons emitting
in the energy range of hard X-rays to soft gamma-rays through
bremsstrahlung. Another mechanism for the production of radio-emitting
electrons is then necessary to explain the radio halo in Coma.  This
conclusion is somewhat in contradiction with the results of Dogiel
(2000) who showed that in this case strong turbulence can accelerate
particles up to energies of $\sim 10$ GeV, albeit with a much flatter
spectrum than needed to produce the observed radio spectrum. The
reason for this disagreement is that Dogiel (2000) used the density
for central regions of the cluster, where the frequency of thermal
particle collisions is high. This led to a higher acceleration
efficiency than in our present treatment which uses average cluster
parameters.

It is shown in Fig.~\ref{loss} that for $q<2$, the synchrotron and IC
losses win over the acceleration term at energies even lower than
1\,GeV. This explains why Blasi (2000), in his study of stochastic
acceleration in the ICM assuming a Kolmogorov spectrum ($q=5/3$)
concluded that {\it in situ} acceleration could not explain the radio
emission because of the synchrotron and IC loss cutoff.  However, we
have shown that the value $q=4.5$, corresponding to weak turbulence
(Sec. 3.2), is consistent with the radio data (Fig.~\ref{radio}).

In the case of $q>3$, however, the acceleration term always wins over
the loss terms at high energies (Fig.~\ref{loss}), which means a rapid
decrease of the characteristic acceleration time, $\tau_{acc}$, with
particle momentum
\begin{equation}
\tau_{acc}\equiv {\hat p^{2}\over \hat a(\hat p)}={p^{2-q}\over{a_0\nu_0}}
\end{equation}
and a larger $q$ results in a smaller value of the injection energy
 at which processes of acceleration dominate.
The best fit parameters, $a_{0} \sim 0.43$ and $q\sim 4.5$, imply a
cluster average $\tau_{acc}
\sim 7\times 10^{6}$\,yrs at the injection energy ($\sim 200$\,keV),
which means that for $E>200$\,keV, $\tau_{acc}$ is much less than the
cluster lifetime.  Hence, the model is self-consistent in the sense that a
`quasi-stationary' state can easily be reached within a cluster lifetime.

 We note, however, the steep spectrum of turbulence ($q\sim 4.5$)
 required for the radio data may pose a problem for the model, since
 it is known that turbulence with $q>2$ cannot effectively scatter
 particles at pitch angles near $\pi/2$ in the quasi-linear regime.
 Nonlinear processes have to be involved to cause effective
 acceleration(e.g., Berezinskii et al. 1990). Further consideration of
 this point is beyond the scope of our model.

Cluster thermal X-ray and radio structures are similar in morphology
(Deiss et al. 1997, Feretti 1999, Liang et al. 2000,  Govoni et
al. 2001). This would be expected from the {\it in situ} model.  We
can calculate the expected relationship between the radio halo and
X-ray profile, if the synchrotron-emitting relativistic particles are
drawn from the thermal pool through {\it in-situ} acceleration. If the
magnetic field is produced by a dynamo process (e.g. turbulence), then
magnetic energy is in equipartition with the energy of turbulent
motions (Kraichnan \& Nagarajan 1967, De Young 1980). Assuming that
the energy of turbulence is proportional to the thermal energy, then
we have the magnetic energy density in a cluster medium proportional
to the thermal energy density $P_{mag} \propto P_{th}$, so that
\begin{equation}
B \propto n_{e}^{1/2}T_{x}^{1/2}\,.
\label{pth}
\end{equation}

As pointed out by En{\ss}lin et al. (1998), this implies that the radio
halo surface brightness profile should be slightly steeper than the
X-ray profile.  The X-ray surface brightness is given by the
projection of $n_{e}^{2}$, which for a $\beta-$model with $n_{e}$ given
by Eq.~\ref{beta} is
\begin{equation}
I_{x}(r)\propto \left[1+(\frac{r}{r_{0}})^{2}\right]^{-3\beta+1/2}
\end{equation}
for fixed $T_{x}$.
 
Since relativistic particles are accelerated from the thermal pool, 
we can assume that their density  is 
proportional to the thermal particle density $n_{e}$.
Then the radio halo brightness profile is given by
\begin{equation}
I_{1.4\,{\rm GHz}} \propto \int n_{e} B^{1+\alpha} \,dl\,,
\end{equation}
when using Eq.~\ref{pth}, and assuming an isothermal ICM of the form of 
Eq.~\ref{beta}
\begin{equation}
I_{1.4\,{\rm GHz}} \propto 
\left[1+(\frac{r}{r_{0}})^{2}\right]^{-3\beta (3+\alpha)/4+1/2}\,.
\end{equation}
 
If the radio spectral index is exactly $\alpha=1$, then the radio and
X-ray profiles follow each other. In the case of Coma and 1E0657-56,
the spectral index is $\alpha \sim 1.25$ and the radio profile is
slightly steeper than the X-ray profile.  For the clusters studied by
Govoni et al. (2001), the radio and X-ray surface brightness profiles
are very similar in shape, implying $\alpha \sim 1$.

\section{Energy Output of Non-Thermal and Quasi-Thermal Particles} 
In this section, for clarity we need to distinguish particles from the
3 different energy ranges: thermal, quasi-thermal and non-thermal. The
Maxwellian thermal particle spectrum is shaped by Coulomb collisions,
where each thermal particle collides with other background particles,
exchanging energies without changing the shape of the energy
spectrum. The lifetime of the thermal particle distribution is
infinite if we neglect thermal bremsstrahlung emission. On the other
hand, a non-thermal particle spectrum is formed by acceleration
processes. If these particles are injected into a region with no
further acceleration, the particles would lose their energy to the
background particles through Coulomb collisions, thus heating the
background plasma and become thermal particles themselves. The lifetime of
non-thermal particles is equal to that of the individual
particles. Unlike the thermal and non-thermal particles, the
quasi-thermal particles exist only in regions of active particle
acceleration. Like the thermal particles, the quasi-thermal (or
suprathermal) particle distribution is formed by Coulomb collisions,
but unlike the thermal particle distribution it is not in strict
equilibrium (but in quasi-equilibrium) and the quasi-thermal
particle distribution would dissipate like the non-thermal particles
once the acceleration process is switched off. Coulomb collisions
cause a net run-away of thermal particles into regions of acceleration
in the spectrum, resulting in a spectral distortion with an excess
above the thermal equilibrium distribution in the quasi-thermal energy
range. Hence the life-time of quasi-thermal particles is shorter than
the thermal particles, but longer than the non-thermal particles.

The energy output of emitting particles is determined by their energy
losses: ionisation losses (Coulomb collisions with background
particles) for quasi-thermal particles, and inverse Compton
scattering and synchrotron radiation losses for non-thermal electrons. The
necessary energy output, $L_{cr}$, can roughly be estimated from the
observed value of X-ray luminosity $L_x$ as
\begin{equation}
L_{cr}\sim L_x{\tau_x\over\tau_{cr}}\,,
\label{output}
\end{equation}
where $\tau_x$ is the characteristic time of X-ray emission by
parent particles and $\tau_{cr}$ is the life-time of these particles.  

If  Coma cluster emits X-rays by inverse Compton process, then from
Eq.~\ref{output} it follows that
\begin{equation}
L_{cr}^C\geq  L_x^C\sim 4\cdot 10^{43}\,\mbox{erg s$^{-1}$}\,,
\end{equation}
where $L_x^C\sim 4\cdot 10^{43}$\,erg s$^{-1}$ is the luminosity of
hard ($20 - 80$\,keV) X-rays from the Coma halo. The uncertainty in
the estimate $L_{cr}^C$ is determined by the unknown rate of
synchrotron losses, since the magnetic field strength is only known
within fairly wide bounds (a range that spans nearly 2 orders of
magnitude).

In the case of a bremsstrahlung origin of the Coma luminosity produced
by non-thermal particles, as in models of En{\ss}lin et al. (1999) and
Sarazin \& Kempner (2000), the necessary energy output $L_{cr}^C$ is
much larger than $L_x^C$. The reason is that these non-thermal
particles lose their energy by collisions with background electrons
and protons (ionisation losses) and the probability of their
generating a bremsstrahlung photon is very low ($<2.5\times 10^{-4}$
for photons with $E_x< 100$ keV; Valinia and Marshall 1998). Therefore
the necessary energy output in this case is huge
\begin{equation}
L_{cr}^C\sim 10^{48}\mbox{erg s$^{-1}$}\,, \
\label{en_nth}
\end{equation} 
and as pointed out by Petrosian (2001), this makes the model problematic. 

Similar problems arise in the interpretation of the hard X-ray flux
from the galactic ridge. The spectrum of hard X-ray emission from the
galactic ridge also shows an excess above the thermal flux. However,
in this case its origin cannot be interpreted as due to inverse
Compton of relativistic electrons (Skibo et al. 1996), thermal
emission of hot plasma or the integrated contribution of discrete
unresolved source (Tanaka 2002).  The only possibility to explain the
origin of hard X-ray flux from the galactic disk is just
bremsstrahlung radiation of sub-relativistic galactic electrons. If
these electrons are non-thermal the necessary energy output should be
of the order of $10^{43}$ erg s$^{-1}$, i.e. more than can be supplied
by supernovae, the most probable sources of energy in the Galaxy (see
e.g. Valinia et al. 2000, and Dogiel et al. 2002a).  It was shown,
however by Dogiel et al. (2002b) that this energy problem can be
resolved if the emission is produced by quasi-thermal electrons. The
life-time of quasi-thermal electrons is longer than that of
non-thermal electrons.  Therefore, as follows from Eq.~\ref{output},
the necessary energy output of quasi-thermal electrons is smaller than
the energy output of non-thermal electrons. We showed above that in
the case of {\it in-situ} acceleration, unlike the models of En{\ss}lin
et al. (1999) and Sarazin \& Kempner (2000) where non-thermal
electrons with power-law spectrum were responsible for the excess hard
X-ray in the Coma cluster, it is the quasi-thermal electrons with
exponential distribution that give rise to the excess hard X-ray
emission. If our model is correct, the energy output in
quasi-thermal electrons needed to explain the Coma X-ray flux is of
the order of $L^C_{cr}\sim 10^{45}$ erg s$^{-1}$ as follows from
numerical calculations for the parameters $a_0=0.43$ and $q=4.5$,
i.e. much less than the estimate in Eq.~\ref{en_nth}.

The bremsstrahlung X-ray flux can also be generated by sub-relativistic
protons (see Dogiel 2001). Though we do not know which specific
process is responsible for particle acceleration in Coma, we note that
in the case of Fermi acceleration the characteristic acceleration
frequency is proportional to $\alpha_0\sim v/L$ (see Toptigyn (1985))
where $v$ is the particle velocity and $L$ is the particle mean free
path (an average distance between scattering centres). Since
velocities of thermal electrons are $\sqrt{M/m}$ times larger than
velocities of thermal protons, we expect that the rate of acceleration
for electrons $\alpha_0^e\gg\alpha_0^p$, hence background
electrons may be accelerated to sub-relativistic energies more
effectively than protons.

\section{Model of `External' Sources of Acceleration}

As follows from Fig.~\ref{loss} the particle spectrum is controlled by
ionisation losses at low energies and synchrotron$+$IC losses at high
energies, which balance the source injection spectrum and
acceleration.  If we assume no {\it in situ} acceleration, i.e. ${\cal
A}(p)=0$ in Eq.~\ref{eq_insitu}, then the kinetic equation for the
distribution function $f$ can be written as:
\begin{equation}
\frac{\partial f(p)}{\partial
t}-\frac{1}{p^{2}}\frac{\partial}{\partial p}{\cal B}(p) f(p)=Q(p) \,.
\end{equation}
The stationary solution of this equation is given by
\begin{equation}
f(p)={1\over{\mid {\cal B}(p)\mid}}\int p^2 Q(p) dp
\end{equation}
i.e. 
\begin{equation}
f(p)={K\over{\mid {\cal B}(p) \mid}(\eta-1)}p^{3-\eta}
\end{equation}
where we assume a power law injection spectrum $Q(p)=K p^{-\eta}$, or
$Q(E) \propto E^{-\gamma_{0}}$ (where $\gamma_{0}=\eta-2$ as follows
from Eq.~\ref{en}).

\begin{figure*}
\epsfxsize 300pt \epsfbox{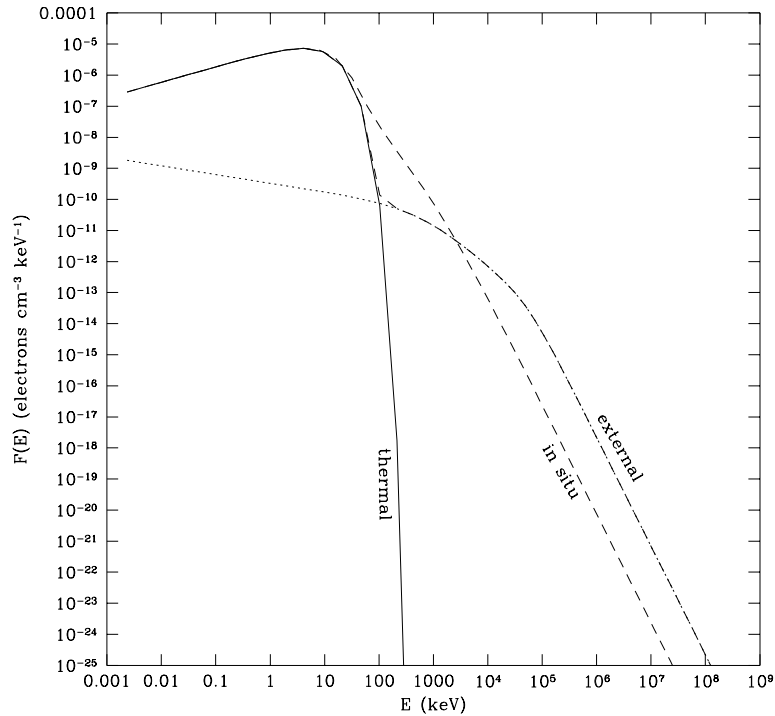}
\caption{The energy distribution of electron density for a pure thermal
distribution is given by the solid curve, and the dotted curve gives
best fit to the X-ray (dominated by IC) and radio data for the
`external' source model. The dot-dashed curve gives the combined
spectrum including the thermal spectrum.  For comparison, the best fit
{\it in situ} model is shown as well (dashed curve). For the
`external' model, it is the electrons at $\sim 5\times 10^{6}$\,keV
that are responsible for the hard X-ray (c.f. the {\it in situ} model
where the hard X-ray is produced by bremsstrahlung of the electrons at
$\sim 50$\,keV).  The cluster is assumed to be isothermal with
$T_{x}=8.2$\,keV and an average electron density  $n_{e} \sim
1.23\times 10^{-4}$\,cm$^{-3}$ within the {\it Beppo-SAX} field of view .}
\label{external}
\end{figure*}
\begin{figure*}
\epsfxsize 300pt \epsfbox{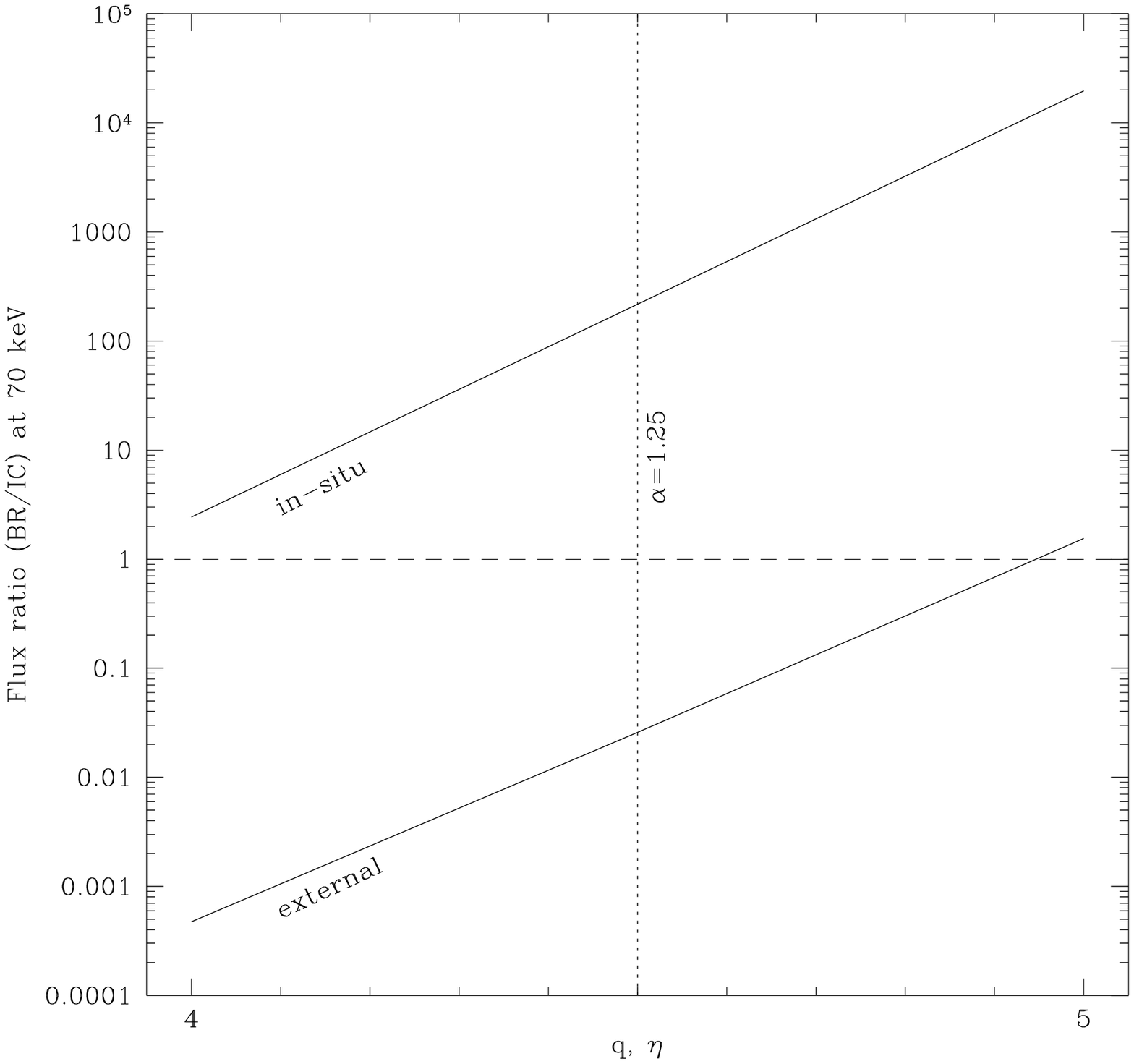}
\caption{The ratio of non-thermal bremsstrahlung to inverse-Compton
emission at 70\,keV derived from the electron energy distributions for
{\it in situ} acceleration at rate $a_{0}p^{-q}$ or the `external'
source model with injection rate $Kp^{-\eta}$. The $q$, $\eta$ value
corresponding to the radio halo spectral index is shown as a dotted
line. The horizontal dashed line marks when the contributions from
non-thermal bremsstrahlung and IC are equal.  }
\label{bric}
\end{figure*}

The best fit to the radio spectrum gives $\eta\sim 4.5$, or $\gamma_{0}\sim 2.5$ (Figure~\ref{external}), just as is
expected for the model of particle acceleration by strong shocks
(Biermann et al. 1993). In this model, the origin of
the hard X-ray emission is definitely IC with no significant
bremsstrahlung (Fig.~\ref{bric}). The model reproduces the
emission spectrum from hard X-rays to radio if the strength of the
magnetic field is less than $0.1-0.2~\mu$G. 
Thus the finding from Faraday rotation measurements (e.g. Feretti et
al. 1995; Clarke et al. 2001) of cluster fields of a few $\mu$G argues
against the IC model for the hard X-ray (e.g. En{\ss}lin et al. 1999
and Dogiel 2000).  However, this is not necessarily a problem.  First,
since there are only a handful of radio sources behind the cluster
with sufficient polarised flux to be detected, the regions of a
cluster probed by these sources may not be representative of the
average field of the entire cluster. Second, since as many as half of
the polarised radio sources used to deduce the cluster magnetic field
are cluster members rather than background sources, we expect that
such measurements would be biased, as a higher $B$-field is likely
near a radio source because of its perturbing effect on the
ICM. Third, Feretti et al. (1995) in their study of the tangling scale
of the magnetic field in the Coma cluster, using an extended polarised
radio source in the centre of the cluster, suggested that the cluster
magnetic field in Coma has a small scale ($<1$\,kpc) component of
strength $\la 5.9h_{50}^{1/2} \mu$G as well as a uniform large scale
($\sim 200$\,kpc) component of strength $0.1-0.3\,\mu$G which is
consistent with the magnetic field of $\sim 0.15\,\mu$G deduced from
the IC model.

Figure~\ref{bric} shows that for this model IC emission dominates
bremsstrahlung at 70\,keV for the required steep spectral index,
$\eta\sim 4.5$.  The presence of strong shocks in galaxy clusters shown
by spatial and temperature images from {\it ROSAT}, {\it ASCA} and now
{\it XMM-Newton} and {\it Chandra} (see Donnelly et al. 1999,
Markevitch et al. 1999, Arnaud et al. 2001, Sun et al. 2001 and Boute
2001) lends observational support to this model.

\section{Conclusions}
We have shown that it is possible to explain both the excess hard
X-ray data and the radio halo spectrum with a model where the
relativistic electrons are weakly accelerated from the thermal pool by
turbulence in the ICM. By using a quasi-stationary solution of the
kinetic equation, we arrive at an electron energy distribution that
deviates from a Maxwellian at $\sim 30$\,keV and reaches a power law
at $\sim 1-10$\,MeV. In the process of reaching an equilibrium, the
collisions between thermal particles produce the
supra-thermal electrons at $\sim 30-100$\,keV required to give the
excess hard X-ray emission through bremsstrahlung emission. The
accelerated GeV electrons are then sufficient to produce the observed
radio halo spectrum with a magnetic field close to the equipartition
value. The model produces a self-consistent electron energy
distribution and has only two free parameters, the acceleration
parameter $a_{0}$ and the momentum spectral index $q$, which are well
constrained by the observed excess hard X-ray spectrum and the radio
halo spectral index (Fig.~\ref{aq}). In the hard X-ray energy range,
the contribution from the inverse Compton scattering of the GeV
electrons is negligible compared with the non-thermal bremsstrahlung
emission for steep radio spectral indices $\alpha > 1$.

It is also possible to explain the hard X-ray and radio
spectrum by a model where accelerated particles are externally
injected in numerous localised regions distributed throughout the
cluster. In this case, the contribution from IC scattering dominates
that of non-thermal bremsstrahlung in the hard X-ray energy range for
steep radio spectral indices.

For both the above models a strong assumption was made, that the
non-thermal X-ray and radio emission from Coma is generated from a
single electron acceleration mechanism. An advantage of this
assumption is that it limits significantly the number of free
parameters in the model. Of course it is possible to have both {\it in
situ} acceleration and external injection of accelerated particles (or
global and local acceleration) at work in the Coma cluster. We have,
however, examined the two extremes rather than such a hybrid
model. This study has shown that if it is purely {\it in situ}
acceleration that produces the particles necessary for the non-thermal
emission in the hard X-ray and radio spectral bands, then the only
explanation for the origin of the hard X-ray is bremsstrahlung of
supra-thermal electrons; and consequently, to fit both X-ray and radio
emission a cluster-wide magnetic field of order $\mu$G (i.e. close to
the equipartition value) is implied. On the other hand, if the
accelerated particles are purely provided by numerous, localised
external sources of injection, then the only possible origin for the
hard X-rays is inverse Compton scattering of CMB photons by
relativistic electrons, which implies a cluster-wide magnetic field of
0.1--0.2\,$\mu$G. Any independent means of measuring the global
cluster magnetic fields would distinguish between the two models.

At present it is not easy to choose between the models of particle
acceleration for Coma. Both models: {\it in situ} acceleration
(global) and shock wave acceleration (local) have their advantages and
problems. In favour of the {\it in situ} model is the similar
morphology between the thermal X-ray and the radio emission. The radio
data have not shown  significant structure in the radio
intensity in the Coma halo which would indicate regions of electron
acceleration. Analysis of the thermal X-ray emission from the Coma
cluster also did not show the strong temperature variations expected
from shocks (Arnaud et al. 2001). On the other hand, several features
indicating recent and ongoing mergers are found in the Coma cluster
(e.g. Donnelly et al. 1999, Arnaud et al. 2001). These mergers often
produce multiple large scale shocks and turbulence capable of both
magnetic field amplification and {\it in situ} electron
acceleration (see a recent paper by Ohno et al. 2002 on the subject). Whether or not these regions extend far enough to allow
electrons to fill the whole volume of the halo, remains to be
discovered.  If the origins of the radio and X-ray emitting electrons
are different, then a much wider range of possible models can be
accommodated. For example, there has been revived interest in the
class of secondary electron models where relativistic and thermal
protons in the ICM collide to produce relativistic (secondary)
electrons (Dennison 1980; Blasi \& Colafrancesco 1999; Dolag \& En{\ss}lin 
 2000; and Miniati et al. 2001).

We would like to thank R. Fusco-Femiano for providing us with the data
points for the PDS spectrum for Coma. We are grateful to the referee
T. En{\ss}lin for helpful comments and suggestions.  VAD thanks the
University of Bristol and the Institute of Space and Astronautical
Science (Sagamihara, Japan) in particular Prof. H. Inoue for
hospitality during his visits.
        
\newpage

\bsp
 
\label{lastpage}


\begin{thebibliography}{}
\bibitem[\protect\citename{Arnaud et al. }2001]{arnaud01}
Arnaud M., Aghanim N., Gastaud R., Neumann D. M., Lumb D., Briel U., Altieri B., Ghizzardi S., Mittaz J., Sasseen T. P., Vestrand W. T., 2001, A\& A, 365, L67
\bibitem[\protect\citename{Berezinskii et al. }1990]{bbdg90}
Berezinskii V. S., Bulanov S. V., Dogiel V. A., Ginzburg V. L., Ptuskin V. S., 1990, Astrophysics of Cosmic Rays, North-Holland.
\bibitem[]{} 
Biermann P. L., Strom R. G., 1993, A\& A, 275, 659
\bibitem[]{}
Blasi P., 2000, ApJ, 532, L9
\bibitem[]{}
Blasi P., Colafrancesco S., 1999, APh, 12, 169
\bibitem[\protect\citename{Blumenthal \& Gould }1970]{bg70}
Blumenthal G. R. \& Gould R. J., 1970, Rev. Mod. Phys., 42, 237
\bibitem[]{}
Bowyer S., Bergh\"ofer T. W., Korpela E J., 1999, ApJ, 526, 592
\bibitem[\protect\citename{Briel et al. }1992]{briel92}
Briel U. G., Henry J. P., B\"ohringer H., 1992, A\& A, 259, L31
\bibitem[]{}
Brunetti G., Setti G., Feretti L., Giovannini, G., 2001, MNRAS, 320, 365
\bibitem[]{}
Bulanov S. V., Dogiel V. A., 1979, Sov. Astron. Lett, 5, 278
\bibitem[]{}
Buote D. A., 2001, astro-ph/0104211
\bibitem[\protect\citename{Bykov \& Toptygin }1990]{bt90}
Bykov A. M., Toptygin I. N., 1993, Physics Uspekhi, 36, 1020
\bibitem[\protect\citename{Bykov et al. }2000a]{bt00a}
Bykov A. M., Chevalier R. A., Ellison D. C., Uvarov Yu. A., 2000a, 
ApJ, 538, 203
\bibitem[\protect\citename{Bykov etal. }2000b]{bt00b}
Bykov A. M., Bloemen H., Uvarov Yu. A., 2000b, A\& A, 362, 886
\bibitem[\protect\citename{Cavaliere \& Fusco-Famiano }1978]{cf78}
Cavaliere A., Fusco-Femiano R., 1978, A\& A, 70, 677
\bibitem[]{}
Clarke T. E., Kronberg P. P., B\"ohringer H., 2001, ApJ, 547, 111
\bibitem[\protect\citename{Cooke et al. }1978]{cooke78}
Cooke  D. A., Lawrence A., Perola G. C., 1978, MNRAS, 182, 661
\bibitem[\protect\citename{Deiss et al. }1997]{deiss97}
Deiss B. M., Reich W., Lesch H., Wielebinski R., 1997, A\& A, 321, 55
\bibitem[]{} 
Dennison B., 1980, ApJ, 239, L93
\bibitem[]{}
De Young, D.S., 1980, ApJ, 241, 81
\bibitem[]{}
Dogiel V. A., Gurevich A. V., Istomin Ia. N., Zybin K. P., 1987, MNRAS, 228, 843
\bibitem[\protect\citename{Dogiel }2000]{dogiel00}
Dogiel V. A., 2000, A\& A, 357, 66
\bibitem[]{}
Dogiel, V.A. 2001, Proc. 4th Integral
Workshop "Exploring the Gamma-Ray Universe" (ESA SP-459), 139
\bibitem[]{}
Dogiel, V.A., Sch\"onfelder, V., Strong, A.W. 2002a, A\&A, 382, 730
\bibitem[]{}
Dogiel, V.A., Inoue, H., Masai, K., Sch\"onfelder, V., Strong, A.W. 2002b, ApJ, submitted
\bibitem[]{}
Dolag, K., En{\ss}lin, T. A., 2000, A\& A, 362, 151
\bibitem[\protect\citename{Donnelly }1999]{don99}
Donnelly R. H., Markevitch M., Forman W., Jones C., Churazov E., Gilfanov M., 
1999, ApJ, 513, 690
\bibitem[\protect\citename{En{\ss}lin \& Biermann }1998]{eb98}
En{\ss}lin T. A., Biermann P. L., 1998, A\& A, 330, 90
\bibitem[\protect\citename{En{\ss}lin et al. }1999]{elb99}
En{\ss}lin T. A., Lieu R., Bierman P. L., 1999, A\& A, 344, 409
\bibitem[]{}
Evrard A. E., Metzler C. A., Navarro J. F., 1996, ApJ, 469, 494
\bibitem[\protect\citename{Felten and Morrrison }1966]{felten66}
Felten J. E., Morrisin P., 1966, ApJ, 146, 686
\bibitem[]{}
Feretti L., Dallacasa D., Giovannini G., Tagliani A., 1995, A\& A, 302, 680
\bibitem[\protect\citename{Feretti \& Giovannini }1996]{fg96}
Feretti L., Giovannini G., 1996, in Ekers R. D., Fanti C., Padrielli L., eds, Proc IAU Symp 175, Extragalactic Radio Sources. Kluwer, p333
\bibitem[]{}
Feretti L., 1999, in H. Boehringer, L. Feretti, P. Schuecker eds., Proc. of the Ringberg workshop, Diffuse Thermal and Relativistic Plasma in Galaxy
Clusters,  Garching: Max-Planck-I
nstitut, p3
\bibitem[\protect\citename{Fusco-Femiano et al. }1999]{fusco99}
Fusco-Femiano R., Dal Fiume D., Feretti L., Giovannini G., Grandi P., Matt G., Molendi S., Santangelo A., 1999, ApJ, 513, L21
\bibitem[\protect\citename{Fusco-Femiano et al. }2000]{fusco00}
Fusco-Femiano R. et al., 2000, ApJ, 534, L7
\bibitem[\protect\citename{Ginzburg \& Syrovatskii }1964]{gs64}
Ginzburg V. L., Syrovatskii S. I., 1964, The origin of cosmic rays. Pergamon press.
\bibitem[\protect\citename{Giovannini et al. }1993]{giov93}
Giovannini G., Feretti L., Venturi T., Kim K.-T. \& Kronberg P. P.,1993, ApJ, 406, 339
\bibitem[\protect\citename{Giovannini et al. }1999]{giov99}
Giovannini G., Tordi M., Feretti L., 1999, New Astronomy, 4, 141
\bibitem[\protect\citename{Giovannini \& Feretti }2000]{giov00}
Giovannini G., Feretti L., 2000, New Astronomy, 5, 335
\bibitem[]{}
Govoni F., En{\ss}lin T. A., Feretti L., Giovannini G., 2001, A\& A, 369, 441
\bibitem[\protect\citename{Gurevich }1960]{gur60}
Gurevich A. V., 1960, Sov. Phys. JETP, 38, 1150
\bibitem[\protect\citename{Hayakawa }1969]{hay69}
Hayakawa S., 1969, {\em Cosmic Ray Physics}, Wiley.
\bibitem[\protect\citename{Herbig \& Birkinshaw }1994]{hb94}
Herbig T., Birkinshaw M., 1994, AAS 185, 3307
\bibitem[\protect\citename{Hughes et al. }1993]{hughes93}
Hughes J. P., Butcher J. A., Stewart G. C., Tanaka Y., 1993, ApJ, 404, 611
\bibitem[]{} 
Jaffe W., 1977, ApJ, 212, 1
\bibitem[\protect\citename{Kempner \& Sarazin }2000]{ks00}
Kempner J. C., Sarazin C. L., 2000, ApJ, 548, 639
\bibitem[]{} 
Kim K.-T., Kronberg P. P., Dewdney P. E., Landecker T. L., 1990, ApJ, 355, 29
\bibitem[]{}
Kraichnan, R.H., Nagarajan, S., 1967, Phys. Fluids, 10, 859
\bibitem[\protect\citename{Liang et al. }2000]{liang00} 
Liang H., Hunstead R., Birkinshaw M., Andreani P.,2000, ApJ, 544, 686
\bibitem[]{}
Liang H., 2000, in Highlights in Astronomy, IAU JD 10, Feretti L. eds. (astro-ph/0012166)
\bibitem[\protect\citename{Markevitch et al. }1999]{mark99}
Markevitch M., Sarazin C., Vikhlinin A., 1999, ApJ, 521, 526
\bibitem[]{}
Miniati F., Jones T. W., Kang H., Ryu D., 2001, ApJ, 562, 233
\bibitem[\protect\citename{Mohr et al. }2000]{mohr00}
Mohr J. J., Reese E. D., Ellingson E., Lewis A. D., Evrard A. E., 2000, ApJ, 544, 109
\bibitem[]{}
Ohno H., Takizawa M., Shibata S., 2002, astro-ph/0206269
\bibitem[]{}
Petrosian V. 2001, ApJ, 557, 560  
\bibitem{} 
Raymond, J. C., Smith, B. W., 1977, ApJS, 35, 419
\bibitem[\protect\citename{Rephaeli }1979]{rephaeli79}
Rephaeli Y., 1979, ApJ, 227,364
\bibitem[]{} 
Roland J, 1981, A\& A, 93, 407
\bibitem[]{}
Sarazin C. L., Kempner J. C., 2000, ApJ, 533, 73
\bibitem[\protect\citename{Schlickeiser }1987]{schlick87}
Schlickeiser R., Sievers A., Thieman H., 1987, A\& A, 182, 21
\bibitem[]{}
Skibo, J.G., Ramaty, R., Purcell, W.R. 1996, A\&AS, 120, 403
\bibitem[]{}
Sun M., Murray S. S., Markevitch M., Vikhlinin A., 2002, ApJ, 565, 867
\bibitem[]{}
Tanaka, Y. 2002, A\&A, 382, 1052 
\bibitem[\protect\citename{Toptygin }1985]{top85}
Toptygin I. N., 1985, Cosmic Rays in Interplanetary Magnetic Fields. Reidel, Amsterdam
\bibitem[]{} 
Tribble P., 1993, MNRAS, 263, 31
\bibitem[]{}
Valinia, A., Marshall, F.E. 1998, ApJ, 505, 134
\bibitem[]{}
Valinia, A., Kinzer, R.L., Marshall, F.E. 2000a, ApJ, 534, 277
\bibitem[\protect\citename{White }2000]{white00}
White D. A., 2000, MNRAS, 312, 663
\end{thebibliography}
\end{document}